\documentclass{wps}

\begin{document}

\markboth{De-ming Wei, Feng-lan Shao, Jun Song, Yun-fei Wang}
{Centrality, system size and energy dependences of charged-particle
pseudo-rapidity distribution}

%
\catchline{}{}{}{}{}
%

\title{Centrality, system size and energy
dependences of charged-particle  pseudo-rapidity
distribution
}

\author{De-ming Wei,$^1$   Feng-lan Shao,$^2$  Jun Song $^2$ and Yun-fei Wang$^2$}

\address{1 Shandong Province Linqu Experimental Middle  School,\\ Shandong
        262600, People's Republic of China;\\ dmwqf2008@hotmail.com}

\address{2 Department of Physics, Qufu Normal University,\\ Shandong
273165, People's Republic of China }
 \maketitle

\begin{abstract}
Utilizing the three-fireball picture within the quark combination
model, we study systematically the charged particle pseudorapidity
distributions in both Au+Au and Cu+Cu collision systems as a
function of collision centrality and energy, $\sqrt{s_{NN}}=$ 19.6,
62.4, 130 and 200  GeV, in full pseudorapidity range. We find that:
(i)the contribution from leading particles to $dN_{ch}/d\eta$
distributions increases with the decrease of the collision
centrality and energy respectively; (ii)the number of the leading
particles is almost independent of the collision energy, but it does
depend on the nucleon participants $N_{part}$; (iii)if Cu+Cu and
Au+Au collisions at the same collision energy are selected to have
the same $N_{part}$, the resulting of charged particle $dN/d\eta$
distributions are nearly identical, both in the mid-rapidity
particle density and the width of the distribution. This is true for
both 62.4 GeV and 200 GeV data. (iv)the limiting fragmentation
phenomenon is reproduced. (iiv) we predict the total multiplicity
and pseudorapidity distribution for the charged particles in Pb+Pb
collisions at $\sqrt{s_{NN}}= 5.5$ TeV. Finally, we give a
qualitative analysis of the $N_{ch}/<N_{part}/2>$ and
$dN_{ch}/d\eta/<N_{part}/2>|_{\eta\approx0}$ as function of
$\sqrt{s_{NN}}$ and $N_{part}$ from RHIC to LHC.
\end{abstract}

\ccode{PACS numbers: 25.75.-q,25.75.Ag}

\section{Introduction}

In relativistic heavy ion collisions at RHIC energies, the charged
particles are produced copiously in vacuum. The number of charged
particles per unit pseudo-rapidity $dN_{ch}/d \eta$, and in
particular its dependence on some variables, such as rapidity,
collision centrality and energy, are  the important observables,
from which a lot of information about the hot and dense matter
created in collisions can be extracted\cite{Xin-Nian
Wang:2001a}\cdash\cite{Eskola:2002qz}. From the pseudo-rapidity
density and the transverse energy per particle, one can determine
via Bjorken method the initial energy density of the fireball which
can provide one piece of evidence for the deconfinement phase
transition. In the fragmentation region, the charged particle
production, in general, is thought to be distinct from that at
mid-rapidity, although there is no obvious evidence for two separate
regions at any of the RHIC energies. The pseudo-rapidity density
$dN_{ch}/d \eta$ in forward rapidity region carries some information
of leading particles produced in collisions \cite{Liu fu-hu}. The
experimental data about the charged-particle pseudo-rapidity density
in both Au+Au and Cu+Cu collision systems have been presented by the
PHOBOS collaboration \cite{Back:2000gw}\cdash\cite{Roland G}, the
PHENIX collaboration \cite{Adcox:2000sp}, and the BRAHMS
Collaboration \cite{Bearden:2001xw,Bearden:2001qq}. The data for the
scaled and shifted pseudo-rapidity distribution
$dN_{ch}/d\eta'/\langle N_{part}/2 \rangle$, exhibit the limiting
fragmentation phenomenon in both Au+Au and Cu+Cu collisions at
different energies and centralities\cite{R.Nouicer,Rachid Nouicer}.

 Recombination of partons \cite{R. J. Fries}\cdash\cite{R. L. Thews},
 Partonic coalescence \cite{Zi-wei Lin}\cdash\cite{Peter F. Kolb} and
 QCM \cite{Feng-Lan Shao:2007a}\cdash\cite{Yao:2006fk}
 have been made to described many observations.
 In our previous work \cite{Feng-Lan
Shao:2007a}, using a Gaussian-like shape rapidity distribution for
constituent quarks as a result of the Landau hydrodynamic
evolution\cite{Edward K.G. Sarkisyn,Edward K.G. Sarkisyn2}, we have
presented the pseudo-rapidity distributions of charged particles in
Au+Au collisions as a function of collision centrality and energy.
The calculation results are in good agrement with the data in
central collisions. In peripheral collisions, our predictions are
slightly lower than data in high rapidity range. The reason may be
that we have not considered the contribution of leading particles.
In present work, taking into account the leading particle influence,
we apply a three-fireball picture \cite{Liu fu-hu,xie:1987a,Liu
Lian-shou} to describe the evolution of the hot and dense quark
matter produced in collisions, and obtain the rapidity distribution
of the constituent quarks just before hadronization. Then let these
constituent quarks combine into initial hadrons according to a quark
combination rule, and allow the resonances in the initial hadrons to
further decay into final hadrons with the help of the event
generator  PYTHIA 6.1 \cite{Sjostrand}.

\section{Three-fireball picture and quark combination model}

In this section, we  introduce the three-fireball picture, which is
used to describe the rapidity distribution of quark and antiquarks
just before hadronization. In addition, we  briefly introduce the
QCM which describes the hadronization of these quarks and antiquarks
produced in collisions.
\subsection{Three-fireball picture}
 It is known that the nucleus-nucleus
collisions at RHIC energies are neither fully stopped nor fully
penetrated. As the incident nuclei penetrate through the target
nuclei, the most of the collision energy is deposited in collision
region to form a big central fireball, and the penetrating quark
matter forms two small fireballs, i.e. target and projectile
fireballs, in forward rapidity region. The charged hadron
pseudorapidity distribution is the total contributions from the
three fireballs.

The big central fireball which contains the main part of collision
energy, controls the rough shape of charged particle pseudorapidity
distribution (width and height). Relativistic hydrodynamics has
successfully described the evolution of system before hadronization.
Here we use a Gaussian-type rapidity distribution for constituent
quarks as a result of the Landau hydrodynamic evolution \cite{Landau
L D}\cdash\cite{Fred Cooper:1973a},

\begin{equation}
f(y)= \frac{1}{\sqrt{2\pi\sigma^2}}\exp
\big(-\frac{y^2}{2\sigma^2}\big) ,\label{eq1}
\end{equation}
where
\begin{equation}
\sigma^2\approx \frac{2c_s^2}{1-c_s^4}\ln
\big(\frac{E\sqrt{s_{NN}}}{2\, m_p \,\epsilon_{c}}\big). \label{eq2}
\end{equation}
Here, $E$ is the effective energy offered by per participant pair,
and it is used to produce the central fireball. $m_p$ is proton mass
and $c_s$ is the sound velocity. $\epsilon_{c}$ is the energy in the
volume of a free hadron at hadronization. All quarks and anti-quarks
in the central fireball are within the rapidity range $[-y_{max},
y_{max}]$,
\begin{equation}
  y_{max}=\frac{c_{s}}{1+c_{s}^2}
\ln{\frac{E\sqrt{s_{NN}}}{2\,m_{p}\,\epsilon_{c}}}.\label{eq3}
\end{equation}

The average constituent quark number in the big central fireball can
be obtained from a simple quark production model \cite{Feng-Lan
Shao:2007a,Xie:1988wi}
\begin{equation}
\label{eq4} \langle{N_q}\rangle=2[(\alpha^{2}+\beta E)^{1/2}-\alpha]
\langle{N_{\rm part}}/2\rangle,
\end{equation}
where the parameter $\beta \approx$ $3.6$ GeV, and the parameter
$\alpha=\beta m-\frac{1}{4}$, $m$ is averaged quark mass and it is
taken to be 0.36 GeV, they are the same with Ref.~\refcite{Feng-Lan
Shao:2007a}.

The two penetrating fireballs mainly consist of leading light
quarks. We also adopt a Gaussian type rapidity distribution for
leading quarks
\begin{equation}
f'(y)=\frac{1}{\langle N_{q(T/P)}\rangle}\frac{dN_{q(T/P)}}{dy} =
 \frac{\exp (-\frac{(y+ y_{0})^2}{2\sigma'^2})}
{\sqrt{2\pi\sigma'^2}}, \label{eq5}
\end{equation}
where  $N_{q(T/P)}$ is the total quark number in the penetrating
target and projectile fireballs. $y_{0}=\pm\frac{ y_{beam}+
y_{max}}{2}$ is the rapidity center for target and projectile
fireballs respectively.  Rapidity distribution range of quarks in
the center of mass frame is $y\in [-y_{beam},-y_{max}]$ for target
fireball and $y\in [y_{max},y_{beam}]$ for projectile fireball,
respectively. In this work, the spectrum width of penetrating
fireballs is taken to be $\sigma' = 0.18$.

The total energy of the three fireballs in nucleus-nucleus
collisions is $\sqrt{s_{NN}}\,\langle N_{part}/2\rangle$,
\begin{equation}
E_{(T+P)}=(\sqrt{s_{NN}}-E)\langle N_{part}/2\rangle,\label{eq6}
\end{equation}
where $E_{(T+P)}$ is total energy of the two penetrating fireballs.
The average number of quarks in penetrating projectile and target
fireballs $\langle N_{q(T+P)}\rangle$ is determined by $E_{(T+P)}$:
\begin{equation}
\langle N_{q(T+P)}\rangle=\frac{E_{(T+P)}}{\langle
E_{q}\rangle}\label{eq7}.
\end{equation}
$\langle E_{q}\rangle$ is the average energy of each quark in the
penetrating projective/target fireballs, and it can be written as:
\begin{equation}
\langle E_{q}\rangle=\int\limits_{y_{max}}^{y_{beam}} {m_{T}\cosh
(y) f'(y)dy}, \label{eq8}
\end{equation}
where $m_{T}=\sqrt{m^{2}+p^{2}_{T}}$ is the transverse mass of
leading quarks.  The transverse momentum $p_T$ of leading quarks is
approximately taken to be 0.25 GeV, one third of the value of
net-proton at forward rapidity $y\approx 3$\cite{Bearden:2004}.
\section{The quark combination model}

The QCM was first proposed for high energy $e^+e^-$ and $pp$
collisions and recently it was extended to ultra-relativistic heavy
ion collisions \cite{Yao:2006fk,Shao:2004}. The model describes the
production of initially produced ground state mesons ($36-plets$)
and baryons ($56-plets$). In principle the model can also be applied
to the production of excited states \cite{Wang:1995ch}. These
hadrons through combination of constituent quarks are then allowed
to decay into the final state hadrons. We take into account the
decay contributions of all resonances of $56-plets$ baryons and
$36-plets$ mesons, and cover all available decay channels by using
the decay program of PYTHIA 6.1 \cite{Sjostrand}. The main idea is
to line up $N_q$ quarks and anti-quarks in a one-dimensional order
in phase space, e.g. in rapidity, and let them combine into initial
hadrons one by one following a combination rule (see section 2 of
Ref.~\refcite{Shao:2004} for a short description of such a rule). We
note that it is very straightforward to define the combination in
one dimensional phase space, but it is highly complicated to do it
in two or three dimensional phase space \cite{Hofmann:1999jx}. The
flavor SU(3) symmetry with strangeness suppression in the yields of
initially produced hadrons is fulfilled in the model
\cite{Xie:1988wi,Wang:1995ch}.
\section{Centrality, system size and energy dependence of charged-particle  pseudo-rapidity distribution}

  In this section, we will study the system size, energy and centrality
dependence of  pseudorapidity distribution of charged particles, and
multiplicity distribution in mid and forward rapidity range
respectively in relativistic heavy ion collisions. Moreover, if the
particle production mechanisms in A+A collisions at RICH and LHC are
the same, we predict the energy and centrality dependences of
$\frac{N_{ch}}{<N_{part}/2>}$ and
$\frac{dN_{ch}}{d\eta<N_{part}/2>}|_{\eta\approx0}$.

From Eq. (\ref{eq6}), we can see that there is only one free
variable, i.e. $E$ or $E_{(T+P)}$,  which should be determined from
the experimental data. In the present work, we determine the
effective energy $E$ for central fireball by fitting the
pseudorapidity density $\frac{dN_{ch}}{d\eta}|_{\eta\approx0}$ in
Au+Au collisions at $\sqrt{s}=130$ GeV. Then the $E_{(T+P)}$ and
leading quark number $\langle N_{q(T+P)}\rangle$ can be naturally
obtained. Using this method, we get the leading quark number in
different collision centralities and parameterize it as the function
of nucleon participants $N_{part}$
\begin{eqnarray}
N_{q(T+P)}=-84.44 + 35.82*N_{part}^{0.4}.\label{eq9}
\end{eqnarray}
At other energies, basing on the relation between the number of
leading quarks and the centrality, we can get $E$, and the
pseudorapidity distribution in different centralities within a quark
combination model.

\begin{figure}
\centerline{\psfig{file=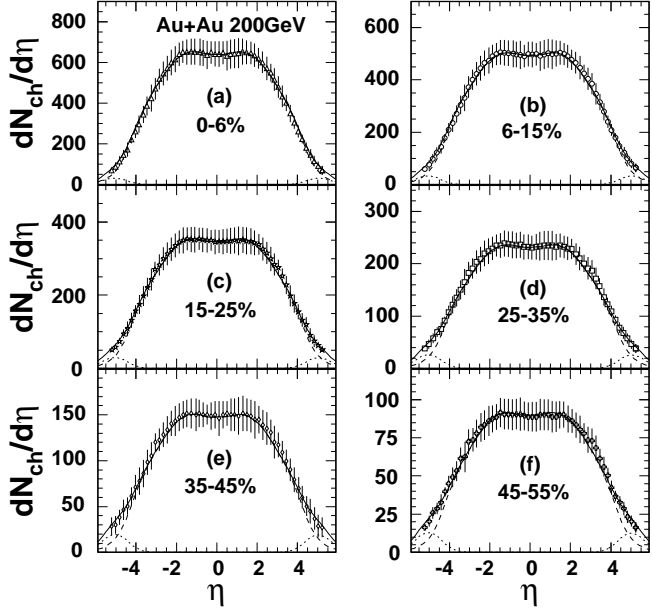,width=10cm}}
 \vspace*{8pt}
\caption{ Charged hadron pseudo-rapidity distributions for different
centralities in Au+Au collisions at $\sqrt{s_{NN}}=200$ GeV. The
lines are our results and the points are data taken from PHOBOS.
}\label{fig1}
\end{figure}
\begin{figure}
\centerline{\psfig{file=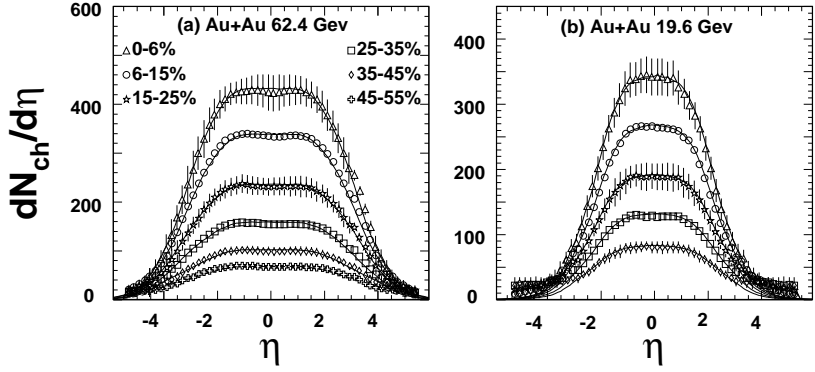,width=12cm}}
 \vspace*{8pt}
\caption{Charged hadron pseudo-rapidity distributions for different
centralities in Au+Au  collisions at $\sqrt{s_{NN}}$=19.6,62.4 GeV.
The lines are our results and the points are data taken from PHOBOS.
}\label{fig2}
\end{figure}
\begin{figure}
\centerline{\psfig{file=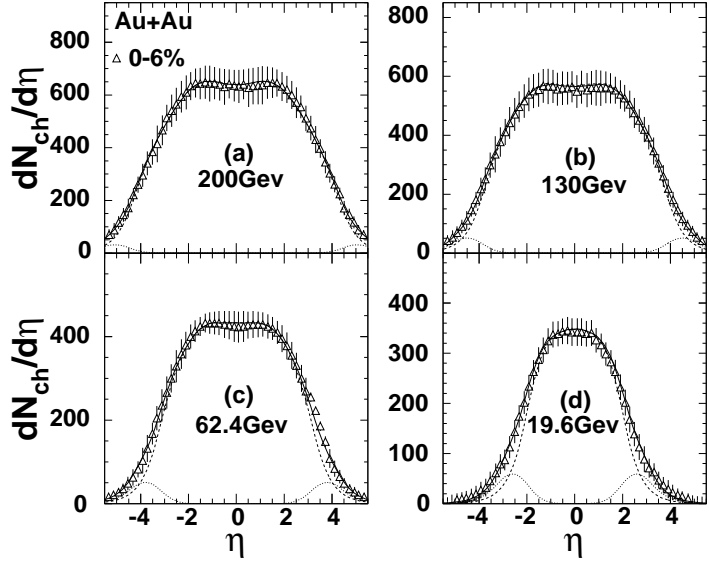,width=10cm}}
 \vspace*{8pt}
\caption{ The energy dependence of charged particles pseudo-rapidity
distributions in most central Au+Au collisions. The lines are our
results. The data are taken from PHOBOS. }\label{fig3}
\end{figure}

Applying Eq.~(\ref{eq9}) to other RHIC energies, we firstly
calculate the charged particle pseudorapidity distributions for
different centralities in Au+Au  collisions at $\sqrt{s_{NN}}=200$
GeV. The data are taken from PHOBOS \cite{Back:2002wb,Back:2002ab}.
The results are shown in Fig.~\ref{fig1}. The dashed lines are the
contribution from the central fireballs. The dotted lines in the
forward pseudorapidity range show the contributions of penetrating
target and projectile fireballs respectively. The solid lines are
the total contribution of the three fireballs. One can see that our
results are in good agreement with the data. In addition, we find
that the contribution from leading particles to $dN_{ch}/d\eta$
distributions increases with the decrease of the collision
centralities. We also give the results of $dN_{ch}/d\eta$
distribution at $\sqrt{s_{NN}}$=19.6, 62.4 GeV, and they are shown
in Fig.~\ref{fig2}. The data are taken from PHOBOS
\cite{Back:2002wb,Back:2002ab}. The agreement of calculated results
with the data is also satisfactory. The Fig.~\ref{fig3} shows the
charged-particle pseudorapidity distributions in most central Au+Au
collisions at $\sqrt{s_{NN}}$=19.6, 62.4, 130 and 200 GeV. The
results indicate that the contribution from leading particles to
$dN_{ch}/d\eta$ distributions increases with the decrease of
collision energy.

\begin{figure}
\centerline{\psfig{file=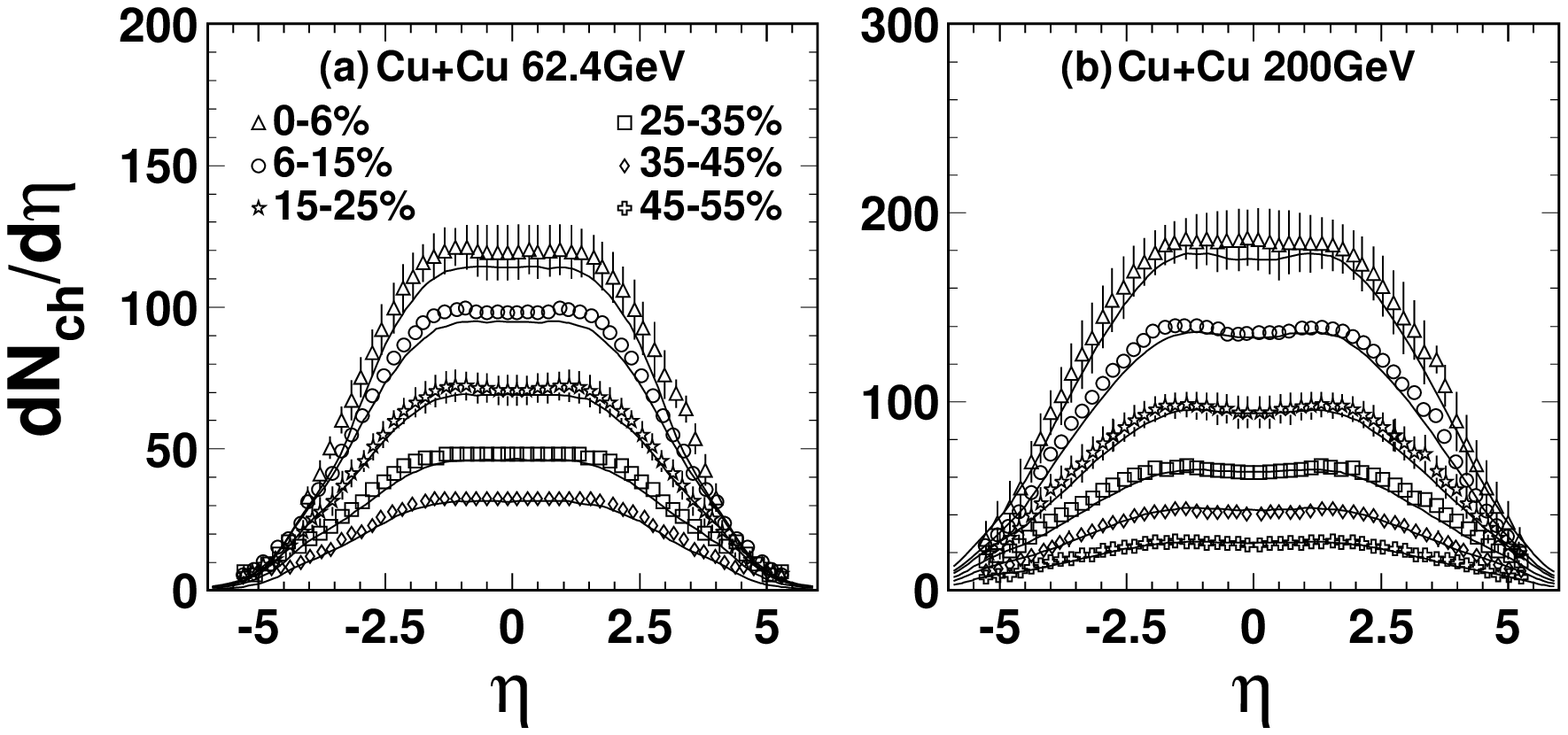,width=13cm}}
 \vspace*{8pt}
\caption{ Particles pseudo-rapidity distributions for different
centralities in Cu+Cu  collisions at $\sqrt{s_{NN}}$=62.4, 200 GeV.
The lines are our results, and the data are taken from PHOBOS.
}\label{fig4}
\end{figure}

Recently, PHOBOS Collaboration have presented the data on
charged-particle pseudorapidity distributions in Cu+Cu collisions at
$\sqrt{s_{NN}}$=62.4, 200 GeV \cite{Roland G}. The other goal of
this paper is to investigate the systematic dependence of particle
production in nuclear collision at RHIC energies, in terms of
overall $dN/d\eta$ distributions. We apply Eq.~(\ref{eq9}) to Cu+Cu
collisions, and give the charged-particle pseudorapidity
distributions as a function of centrality at $\sqrt{s_{NN}}$=62.4,
200 GeV. The results are shown in Fig.~\ref{fig4}, and compared with
the data. As we can see, our results are  roughly consistent with
the experimental data, but are slightly lower than the data in the
mid-rapidity range, especially for the central Cu+Cu collisions. The
reason may be that we did not consider the difference  between the
collision geometry in Cu+Cu and Au+Au, even at the same $N_{part}$
in Eq.~(\ref{eq9}), especially for the central Cu+Cu collisions. The
two system have same shape, which indicate a similarity particle
production mechanism between Au+Au and Cu+Cu, the results is the
same as Ref.~\refcite{Bialas and A. Bzdak}. Following that, we think
the Au+Au and Pb+Pb collisions have the same particle production
mechanism, so we predict the pseudorapidity distribution and charged
particle multiplicity in Pb+Pb collisions at $\sqrt{s_{NN}}$=5.5 TeV
in the followings.
\begin{figure}
\centerline{\psfig{file=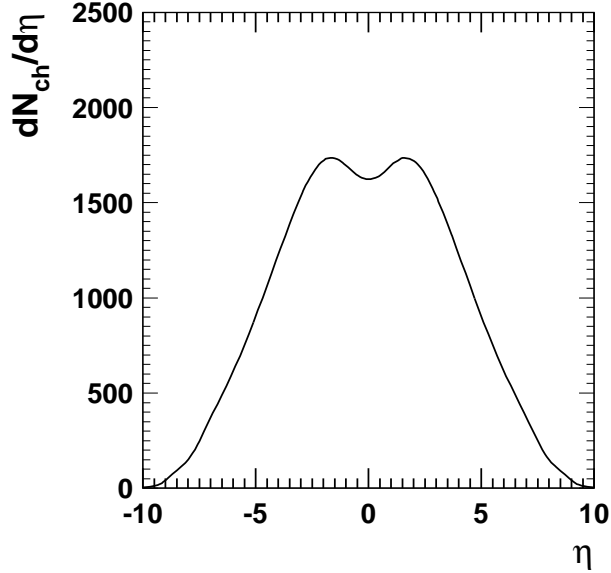,width=10cm}}
 \vspace*{8pt}
\caption{Charged particles pseudorapidity distributions calculated
by QCM in most central Pb+Pb collisions at $\sqrt{s}=5.5$
TeV.}\label{fig5}
\end{figure}
\begin{figure}
\centerline{\psfig{file=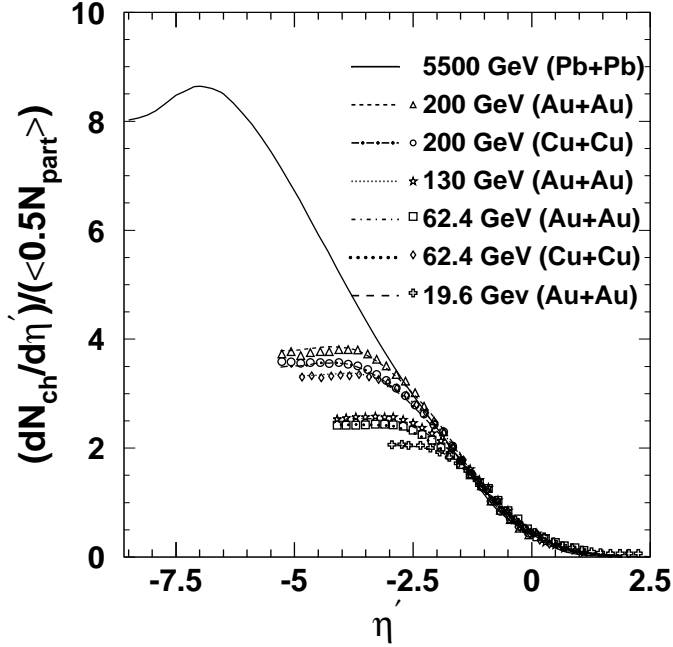,width=10cm}}
 \vspace*{8pt}
\caption{The scaled, shifted pseudorapidity density at
$\sqrt{s_{NN}}$=19.6,62.4,130,200 GeV in most central
Au+Au,Cu+Cu,and Pb+Pb collisions at $\sqrt{s}$=5.5TeV. The lines are
our results, the symbols are  data taken from PHOBOS.}\label{fig6}
\end{figure}
\begin{figure}
\centerline{\psfig{file=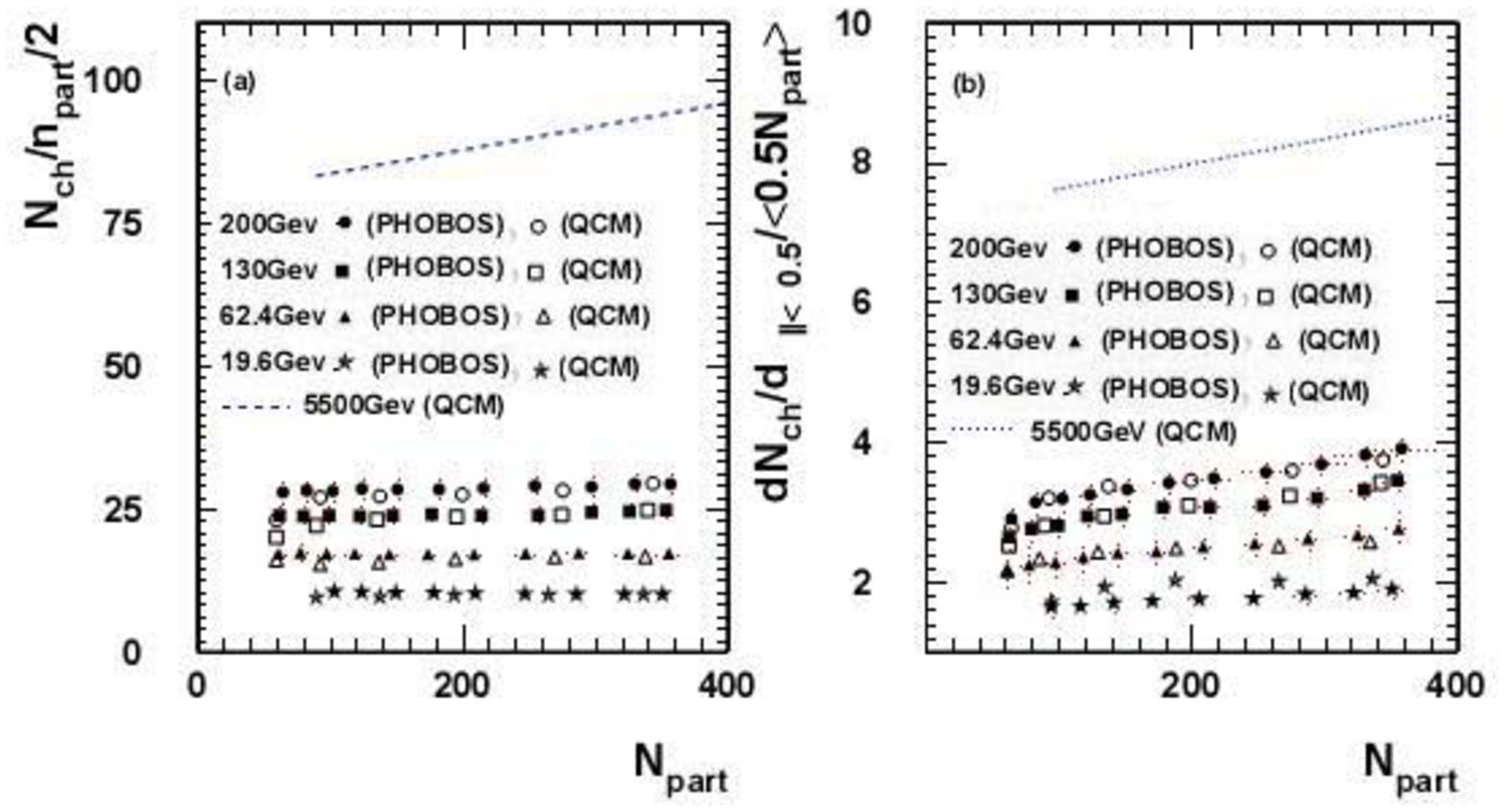,width=13cm}}
 \vspace*{8pt}
\caption{The total number of charged particles (a) and mid-rapidity
density (b) per participant pair shown as a function of $N_{part}$
for $\sqrt{s}$ =19.6, 62.4, 130 and 200 GeV in Au+Au collisions, and
$\sqrt{s}$ =5500 GeV in Pb+Pb collisions. The solid symbols  are
data taken from PHOBOS, and the open symbols are our results in (a)
and (b). The lines are our prediction at LHC in (a) and
(b).}\label{fig7}
\end{figure}

The Large Hadron Collider (LHC) is scheduled to begin operation in
May 2008. The most pressing issue for the early days at the LHC is
to establish the global features of heavy ion collisions. This
involves the estimation of the inclusive charged-particle yield and
the charged pseudorapidity distribution and so on. In this work,
extending Eq.~ (\ref{eq9}) to LHC energy, we can predict the charged
particle pseudorapidity distribution as a function of centrality in
Pb+Pb collisions. As an example, we calculate the most central Pb+Pb
collisions at $\sqrt{s_{NN}}=5500$ GeV in Fig.~\ref{fig5}. The total
charged-particle multiplicity is about $18170$, and pseudorapidity
density $dN_{ch}/d\eta\mid_{\mid\eta\mid<0.5}$ is about 1630.
 Note that the sound velocity  is taken be 1/3. To
separate the trivial kinematic broadening of the $dN_{ch}/d\eta$
distribution from the more interesting dynamics, we also study the
scaled, shifted pseudorapidity distribution $dN_{ch}/d\eta'/\langle
N_{part}/2 \rangle$ , where $\eta'=\eta -y_{beam}$ , in  most
central Au+Au, Cu+Cu and Pb+Pb collisions at different energies. The
calculation results are shown in Fig.~\ref{fig6}. We observe that
the data at various energies and systems fall on a common limiting
curve.

The centrality dependence of $N_{ch}/<N_{part}/2>$ and
$dN_{ch}/d\eta/<N_{part}/2>|_{\eta\approx0}$ at
$\sqrt{s_{NN}}$=19.6, 62.4, 130, 200 GeV in Au+Au collisions (open
symbol) compared with the data (solid symbol) taken from PHOBOS
\cite {Roland G} and the prediction at 5500 GeV in Pb+Pb collisions
(the solid line) are shown in Fig.~\ref{fig7}.

\begin{figure}
\centerline{\psfig{file=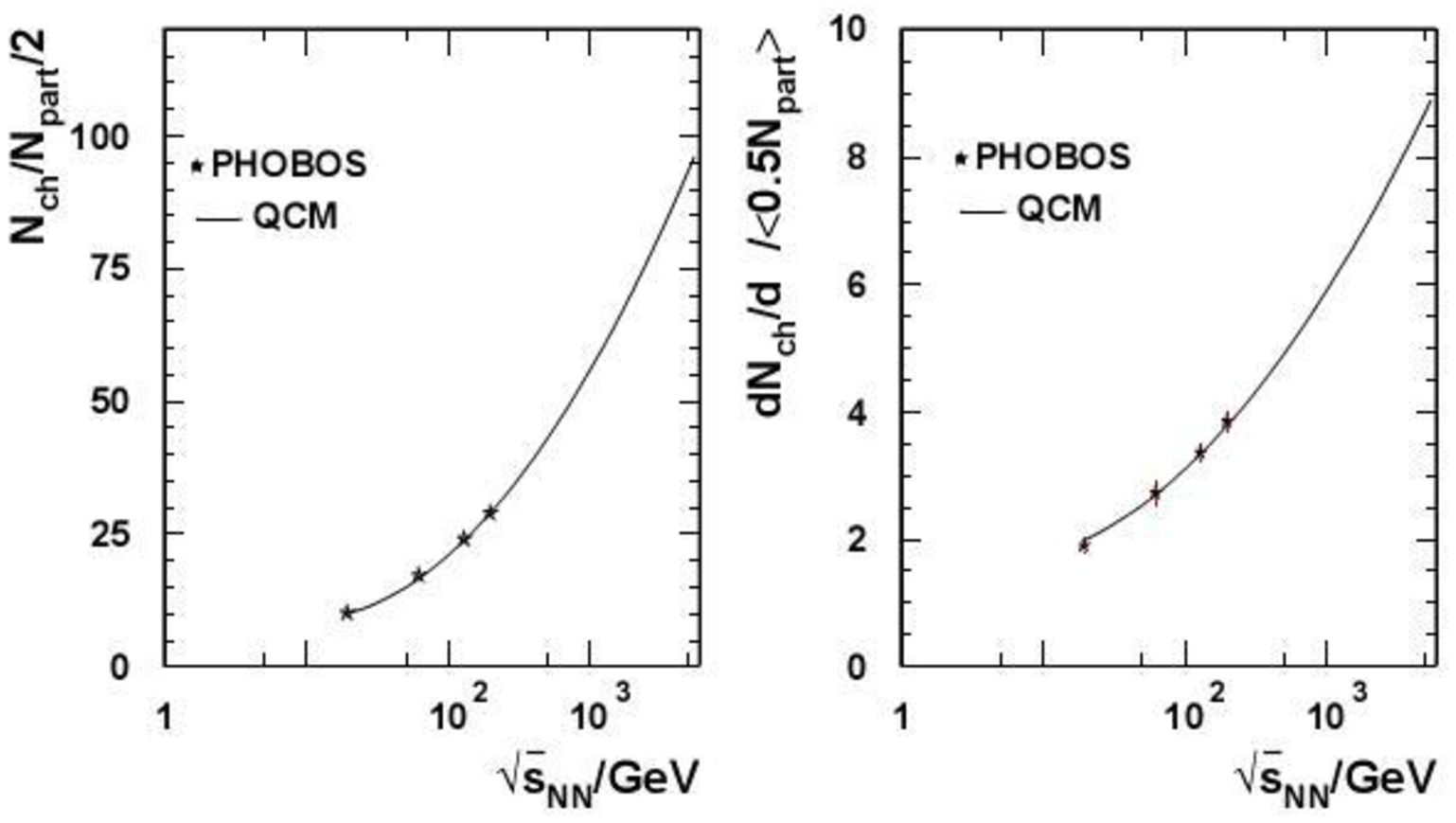,width=12cm}}
 \vspace*{8pt}
\caption{The charged particles mean multiplicity $<N_{ch}>$ and
pseudorapidity density of charged particle per participant pair
($N_{part}/2$ ) as a function of the c.m. energy of collision. The
lines are our results and the symbols are data taken from PHOBOS.
}\label{fig8}
\end{figure}

 Fig.~\ref{fig8} shows the  c.m. energy dependence
of $N_{ch}/<N_{part}/2>$ and
 $dN_{ch}/d\eta/<N_{part}/2>|_{\eta\approx0}$ from RHIC to LHC predicted by the QCM (the lines). It
 shows that the $N_{ch}/<N_{part}/2>$ and
$dN_{ch}/d\eta/<N_{part}/2>|_{\eta\approx0}$ from RHIC to LHC can
grow at logarithmically with $\sqrt{s_{NN}}$.

\section{summery}
Within a combination model, we study the charged particle
pseudo-rapidity distributions in Au+Au and Cu+Cu collision systems
as a function of collision centrality and energy ($\sqrt{s_{NN}}=
19.6, 62.4, 130$ and $200 $ GeV), in full pseudo-rapidity range. We
use a toy model, i.e. three fireballs, to describe the evolution of
the hot and dense quark matter produced in collisions. The big
central fireball which carries the main part of collision energy
controls the rough shape of the charged particle pseudo-rapidity
distribution. We apply the Landau relativistic hydrodynamic model to
describe the the evolution of highly excited and possibly deconfined
quark matter created in the big central fireball. As a result, we
obtain a Gaussian-type rapidity spectra of constituent quarks before
hadronization. The other two small fireballs in foreword rapidity
carry the information of the leading particles. We also use a
Gaussian-type rapidity spectra of constituent quarks before
hadronization. Then we use our combination model to describe the
hadronization of initially produced hadrons including resonances,
whose decays are dealt with by the event generator PYTHIA 6.3
\cite{Sjostrand}.
Firstly, by studying the contribution of leading particles to
charged-particle pseudo-rapidity distribution in Au + Au collisions
for different centralities at $130$ GeV, we extract the centrality
dependence of the average number of leading quarks from the data.
Then we extend it to other RHIC energies. We calculate the charged
particle pseudo-rapidity distributions in both Au+Au and Cu+Cu
collision systems as a function of collision centrality, at
$\sqrt{s_{NN}}=$ 19.6, 62.4 and 200 GeV, in full pseudo-rapidity
range. The calculation results are in good agreement with data. To
separate the trivial kinematic broadening of the distributions of
the pseudo-rapidity density from more interesting dynamics, we
compute the scaled and shifted pseudo-rapidity density distributions
$dN_{ch}/d\eta '/\langle N_{part}/2\rangle$ with $\eta '=\eta -y
_{beam}$ at collision energies 19.6, 62.4, 130 and 200 GeV. The good
agreement with data is found.   Furthermore, we predict the total
multiplicity and pseudo-rapidity distribution for the charged
particles in most central Pb+Pb collisions at $\sqrt{s_{NN}}= 5.5$
TeV. Through investigating detailed  $dN_{ch}/d\eta$ distributions,
we find that: (i)The contribution from leading particles to
$dN_{ch}/d\eta$ distributions increases with the decrease of the
collision centrality and energy respectively; (ii)The number of
leading particles is, independent of collision energy, only a
function of nucleon participants $N_{part}$for the same system;
(iii)If Cu+Cu and Au+Au collisions at the same collision energy are
selected to have the same $N_{part}$, the resulting of charged
particle $dN/d\eta$ distributions are nearly identical, both in the
mid-rapidity particle density and the width of the distribution.
This is true for both 62.4 GeV and 200 GeV data. (iv)The limiting
fragmentation phenomenon is reproduced. Furthermore, we predict the
total multiplicity and pseudorapidity distribution for the charged
particles in Pb+Pb collisions at $\sqrt{s_{NN}}= 5.5$ TeV, and find
the $N_{ch}/<N_{part}/2>$ and
$dN_{ch}/d\eta/<N_{part}/2>|_{\eta\approx0}$ from RHIC to LHC can
grow at logarithmically with $\sqrt{s_{NN}}$.
\section*{Acknowledgments}
The authors thank Qu-bing Xie, Wei-han and Tao-yao for helpful
discussions. The work is supported in part by the National Natural
Science Foundation of China under the grant 10775089 and 10475049.

\end{document}